# QED DERIVED FROM THE TWO-BODY INTERACTION (2)
## *THE DERIVATION*

Sarah B. M. Bell,[1,2] John P. Cullerne,[1] Bernard M. Diaz[1,3]

## Abstract

We have shown in a previous paper that the Dirac bispinor can vary like a four-vector and that Quantum Electrodynamics (QED) can be reproduced with this form of behaviour.[2] We have also shown in part (1)[3] of this paper, that QED with the same transformational behaviour also holds in a second space which we called *M*-space. Here we use *M*-space to show that QED can be reduced to two simple rules for a two-body interaction.

## 1. INTRODUCTION

### 1.1 Summary of part (1)

We have shown in a previous paper that the Dirac equation may be solved to yield a wave function that is a reflector matrix whose quaternion elements are the Dirac bispinors.[2] In this novel representation of a bispinor, it is easy to show that under spatial rotation, Lorentz transformation, charge conjugation, parity change and time reversal, both the bispinor and the mass term can vary like a four-vector. There is also a conserved charge density current where these variables have the same behaviour, which leads to a photon equation. The photon and Dirac equations hold in the usual four-dimensional flat spacetime which we called $L$. In part (1) of this paper[3] we showed that the photon and Dirac equations with the same transformational behaviour also hold in a second four-dimensional spacetime we called $M$ with an alternative topology. $M$ has a scalar parameter, $R$, describing a curve associated with the space, although $M$ itself is flat. We may also include a Lorentz transformation, $Z$, relative to a fixed frame in $L$.

---

[1] Department of Computer Science I.Q. Group, The University of Liverpool, Chadwick Building, Peach Street, Liverpool, L69 7ZF, United Kingdom.

[2] U.K. phone number and email address, 01865 798579 and Sarabell@dial.pipex.com

[3] E-mail, B.M.Diaz@csc.liv.ac.uk

We defined a bijection between points in $L$ and points in $M$ which allowed us to determine the behaviour of a particle in $M$ given its behaviour in $L$ or the reverse. Using this, we discovered how to relate the charge density current in $M$ to that in $L$ which meant we could transfer an interaction in $L$ into $M$ or the reverse.

We described the two-body attractive interaction by assuming one body generates a potential field in conformity with the photon equation while the behaviour of the second body is determined by solving the Dirac equation with this potential. We discovered how to model the attractive two-body interaction in $M$ using the photon and Dirac equations. We showed that the solutions of the Dirac equation resembled the orbits of Bohr.[7, 8] We deduced the same two equations as Bohr did to describe such two-body interactions, although we transferred them into the new space *M*. We called these equations *the Bohr equations*. We called our model of the two-body interaction in *M the Bohr interaction*.

We resolved the discrepancy between the integer angular momentum for the two-body interaction predicted by the Bohr equations for *M* and the half-integer angular momentum found when the Dirac equation is solved in the usual way in *L*, for example for the one-electron atom.[6] We called this latter structure *the Dirac interaction*. We showed that the two methods of describing the two-body interaction, the Bohr and Dirac interaction, lead to similar results. Since the energy and angular momentum eigenvalues are the same for both, we concluded that there is nothing preventing a Dirac interaction from becoming a Bohr interaction or vice versa, except possibly the symmetry of the boundary conditions.

## 1.2   Overview of part (2)

The above emergent structure begs the further step of varying *R* and Z locally. *M* becomes $M_q$ where *q* varies with the location in *L*. This means we can generalise further than a single two-body interaction in *M*. Each $M_q$ can represent a different two-body interaction. We find we can then relate these to the Dirac and photon equation, using the Bohr equations.

We discuss the underlying strategy. The Dirac equation for point *q* tells us in what direction and at what speed a particle would move if it were at location *q*. The equation therefore makes a copy of the particle for each location. The equation also tells us the probability of the particle being at $q$ but this is an unconnected function. The photon equation tells us the potential due to a local charge density current, the current being different at each location. Therefore we need a single particle, which we call $N^b$, to represent the many possible local interactions for the Dirac equation, but many particles, which we call $N_q^a$, to represent the local currents for the photon equation. The probability of particle $N^b$ being at point $q$ is constrained by the boundary conditions and the requirement that the probability should vary smoothly which the Dirac equation imposes. We are modelling the equations in all their generality without global boundary conditions, and so we do not constrain the probability of particle $N^b$ being at point $q$.

We summarise. We consider a Bohr interaction between $N^b$ and a single $N_q^a$. We show that the photon equation is valid on $L$ and also $M_q$ for a suitable set of points, $b_q$, at a distance $R_q$ from point $q$. We then use the Bohr equations to construct a wave function that is necessarily a solution of the Dirac equation in $M_q$ and also in $L$ for the same set of points, $b_q$.

We move from the photon and Dirac equation being satisfied at a single set of points, $b_q$, for one particle $N_q^a$ to many such particles, where $q$ takes multiple values and finally the photon and Dirac equations are valid for all of $L$. We do this by considering two forms of limit behaviour as $R_q \rightarrow 0$. For the first form, we prove the result for a network of the points, $b_q$, at different distances, $R_q$, from point $q$. This method permits discontinuous behaviour, broadening the scope of the photon and Dirac equations. For the second form, we put the equations onto a lattice, and prove the result holds if the variables show the usual continuous behaviour. We find on doing this that the mass of particle $N^b$ shows typical lattice behaviour varying inversely with a power of the lattice spacing. The charge on particle $N^b$ and the sum of the charges on the particles $N_q^a$ remain finite.

### 1.3 Results from the previous paper

The versatile Dirac equation for an electron is

$$\left\{\underline{\mathbf{D}}\left(\mathbf{D}, \mathbf{D}^{\ddagger}\right) - ie\underline{\mathbf{A}}\left(\mathbf{A}^{\sim}, \mathbf{A}^{\sim\ddagger}\right)\right\}\underline{\boldsymbol{\Phi}}(\phi_1, \phi_2) \qquad (1.3.\mathrm{A})$$
$$= \underline{\boldsymbol{\Phi}}(\phi_1, \phi_2)\underline{\mathbf{M}}\left(\mathbf{M}, -\mathbf{M}^{\ddagger}\right)$$

The versatile photon equation is

$$\underline{\mathbf{D}}\,\underline{\mathbf{D}}\,\underline{\mathbf{A}}^{\sim} = \underline{\mathbf{J}}^{\sim}\left(\mathbf{J}^{\sim}, \mathbf{J}^{\sim\ddagger}\right) \qquad (1.3.\mathrm{B})$$

We may simplify this equation to yield

$$\mathbf{D}\mathbf{D}^{\ddagger}A_{\mu} = J_{\mu} \qquad (1.3.\mathrm{C})$$

where $A_{\mu}$ is a component of the potential and $J_{\mu}$ of the Dirac current. The other variables are defined in part (1) and also discussed in more detail by Bell et al.[2]

In part (1) we defined two spaces, spacetime, which we called $L$ and co-ordinated with Cartesian co-ordinates $(x_0, x_1, x_2, x_3)$, and $M$-space which we co-ordinated with $(x_0, s, r, x_3)$. The polar co-ordinates for both $L$ and $M$ are $(r, \theta)$ and for $L$, $s' = r\theta$, where $s'$ is the arc corresponding to $\theta$, while for $M$, $s = R\theta$, where $s$ is the arc corresponding to $\theta$ and $R$ is constant. There is a simple bijection, between the co-ordinates of a point in $M$ and $L$. For each $x_0$, $\theta$, $r$ and $x_3$ are the same and

$$s' = \frac{r}{R}s \qquad (1.3.D)$$

We showed that a potential term of $\mathbf{A}^\sim$ in $L$ corresponds to a potential term of $\mathbf{A}^{M\sim}$ in $M$, where

$$\mathbf{A}^{M\sim} = \frac{\mathbf{A}^\sim r}{R} \qquad (1.3.E)$$

We solved the Dirac equation for the two-body interaction in $M$

$$\left\{ \underline{\mathbf{D}}\left(\mathbf{D}, \mathbf{D}^\ddagger\right) - ie\underline{\mathbf{A}}^{M\sim}\left(\mathbf{A}^{M\sim}, \mathbf{A}^{M\sim\ddagger}\right) \right\} \underline{\boldsymbol{\Phi}}^M\left(\phi_1^M, \phi_2^M\right) \qquad (1.3.F)$$
$$= \underline{\boldsymbol{\Phi}}^M\left(\phi_1^M, \phi_2^M\right) \underline{\mathbf{M}}\left(\mathbf{M}, -\mathbf{M}^\ddagger\right)$$

where $\underline{\boldsymbol{\Phi}}^M$ is the solution associated with the new potential term $\mathbf{A}^{M\sim}$, with

$$\mathbf{A}^{M\sim} = \frac{if}{R} \qquad (1.3.G)$$

and found

$$\phi_1^M = \exp\left\{ i\left(v^\sim x_0^\sim + \mu s\right) \right\}, \qquad (1.3.H)$$
$$\phi_2^M = i\left\{ v^\sim - \mathbf{i}_s\mu - eA^{M\sim} \right\} \mathbf{M}^{-1} \exp\left\{ i\left(v^\sim x_0^\sim + \mu s\right) \right\},$$
$$m^{\sim 2} = \left(v^\sim - eA^{M\sim}\right)^2 + \mu^2$$

where $\mathbf{i}_s$ is the quaternion matrix associated with $s$, $x_0^\sim = x_0/i$, $e$ is the charge on the electron, $m^\sim = m_e/i$ and $m_e$ is the mass, $iv^\sim$ is the frequency and $\mu$ the wave number of the electron. We called this bound state for the two-body attractive interaction in *M*, *the Bohr interaction*.

We quoted the de Broglie relations for the energy and momentum of the Bohr electron

$$\eta^\sim = \frac{m^\sim}{\sqrt{1+v^{\sim 2}}} \qquad (1.3.\text{I})$$

where $i\eta^\sim$ is the frequency and $-iv^\sim = -i\mu/\eta^\sim$ the velocity of the Bohr electron and

$$\mu = \frac{m^\sim v^\sim}{\sqrt{1+v^{\sim 2}}} \qquad (1.3.\text{J})$$

We derived the Bohr equations of which the first is

$$\frac{ief}{R} = \frac{m^\sim v^{\sim 2}}{\sqrt{1+v^{\sim 2}}} \qquad (1.3.\text{K})$$

and the second is

$$\frac{m^\sim v^\sim R}{\sqrt{1+v^{\sim 2}}} = n\frac{h}{2\pi} \qquad (1.3.\text{L})$$

where *n* is an integer.

## 2. QED EQUATIONS FOR A POINT

### 2.1 The photon equation

Suppose we have a sphere of radius $r$, centred at a point *q* in *L*, with a particle $N_q^a$ of charge $f_q$ at rest at $q$. The Bohr equations imply an inverse distance square law for the force of attraction in $L$. In that case, the potential, $iA_q^\sim$, at a point $b_q$ on the boundary of the sphere in the same co-ordinate frame is given by

$$A_q^\sim = \frac{if_q}{r} \qquad (2.1.A)$$

We may replace $f_q$ by a constant charge density $i\rho_q^\sim$. We set the total charge within the sphere to the same value

$$f_q = -i\frac{4\pi r^3 \rho_q^\sim}{3} \qquad (2.1.B)$$

and since the charged sphere produces the same potential at $b_q$ as the isolated charge at the centre, we have from equation (2.1.A)

$$A_q^\sim = \frac{4\pi r^2 \rho_q^\sim}{3} \qquad (2.1.C)$$

The d'Alembertian reduces to

$$\mathbf{DD}^\ddagger = \frac{\partial^2}{\partial r^2} \qquad (2.1.D)$$

and we see from equation (2.1.C)

$$\mathbf{DD}^\ddagger A_q^\sim = \frac{8\pi \rho_q^\sim}{3} \qquad (2.1.E)$$

Equation (2.1.E), apart from a constant indicating a different choice of units, is an alternative form of the photon equation in the rest frame of $N_q^a$ in $L$, as we can see from equation (1.3.C).

Let $r$ be $R_q$, the Bohr radius. The potential term is, from equation (2.1.A),

$$A_q^{\tilde{}} = \frac{if_q}{R_q} \qquad (2.1.F)$$

and from equation (1.3.E) this is also the potential term in the M-space $M_q$. Equation (2.1.C) becomes

$$A_q^{\tilde{}} = \frac{4\pi R_q^2 \rho_q^{\tilde{}}}{3} \qquad (2.1.G)$$

## 2.2  The Dirac equation

We suppose that particle $N^b$ of charge $e$ and mass $im^{\tilde{}}$ positioned at the point $b_q$ in $M_q$ interacts with the particle $N_q^a$ in conformity with the Bohr equations, (1.3.K) and (1.3.L). Using these equations, we may calculate the velocity, $-iv_q^{\tilde{}}$, of particle $N^b$ relative to particle $N_q^a$

$$v_q^{\tilde{}} = i\frac{2\pi e f_q}{nh} \qquad (2.2.A)$$

and a condition on $m^{\tilde{}}$ and $R_q$

$$m^\sim R_q = -i \frac{n^2 h^2 \sqrt{1 - \frac{4\pi^2 e^2 f_q^2}{n^2 h^2}}}{4\pi^2 e f_q} \qquad (2.2.B)$$

We may calculate the appropriate wave number, $\mu_q$, for particle $N^b$ from de Broglie's relation, equation (1.3.J),

$$\mu_q = \frac{m^\sim v_q^\sim}{\sqrt{1 + v_q^{\sim 2}}} \qquad (2.2.C)$$

From our discussion in part (1), we may also calculate the appropriate frequency for the wave function of particle $N^b$ from the energy, $i v_q^\sim$, of the particle, which is the sum of the kinetic energy, given by the de Broglie relation, equation (1.3.I), and potential energy, arising from the electromagnetic potential in equation (2.1.F),

$$v_q^\sim = \frac{m^\sim}{\sqrt{1 + v_q^{\sim 2}}} + i \frac{e f_q}{R_q} \qquad (2.2.D)$$

We may now assemble a wave function, $\underline{\boldsymbol{\Phi}}^{M_q}\big(\phi_1^{M_q}, \phi_2^{M_q}\big)$, in analogy to (1.3.H). This is

$$\phi_1^{M_q} = \exp\{i(v_q^\sim x_0^\sim + \mu_q s)\}, \qquad (2.2.E)$$
$$\phi_2^{M_q} = i\{v_q^\sim - \mathbf{i}_s \mu_q - e A_q^\sim\}\mathbf{M}^{-1} \exp\{i(v_q^\sim x_0^\sim + \mu_q s)\},$$
$$m^{\sim 2} = (v_q^\sim - e A_q^\sim)^2 + \mu_q^2$$

$\mu_q$ and $v_q^\sim$ may be expressed in terms of $e$, $f_q$, $n$, and $m^\sim$ from equations (2.2.C), (2.2.D), (2.2.A) and (2.2.B), where we again set $h/2\pi$ to 1. $A_q^\sim$ may be expressed in the same terms from equations (2.1.F) and (2.2.B). We know that $\underline{\boldsymbol{\Phi}}^{M_q}$ is a solution to the Dirac equation

$$\left\{\underline{\mathbf{D}} - ie\underline{\mathbf{A}}_q^{\sim}\left(A_q^{\sim}, A_q^{\sim}\right)\right\}\underline{\boldsymbol{\Phi}}^{M_q} = \underline{\boldsymbol{\Phi}}^{M_q}\underline{\mathbf{M}}\left(m^{\sim}, -m^{\sim\ddagger}\right) \qquad (2.2.F)$$

in the space $M_q$, where $A_q^{\sim}$ is constant. In the light of our discussion in part (1) of this paper, we also know that the solution of equation (2.2.F) in $M_q$ is accompanied by a solution in $L$. We wish to know the class of points, $\hat{b}_q$, for which the solution, with the same potential term, $A_q^{\sim}$, holds in $L$.

We have two alternatives for where the particle $N^b$ may appear in $L$, other positions being forbidden by the conservation of energy as discussed in part (1). For the first, we have a single state as given in equation (2.2.E), in which case $\hat{b}_q$ lies on the boundary of a circle, centre $q$, radius $R_q$, in the $(x_1, x_2)$ plane in $L$. The spatial part of the wave function in $M_q$ depends only on the variables in a single plane. We call this *a pure state*. For the second alternative we have a superposition of rotations of the state given by equation (2.2.E), and $\hat{b}_q$ may lie anywhere on the surface of the sphere, centre $q$, radius $R_q$, in $L$. The spatial part of the wave function in $M_q$ depends on all the spatial co-ordinates. We call this *a superposition state*. We call the circle or sphere whose boundary is delimited by the set of points $\hat{b}_q$ *a roundel*. We identify $\hat{b}_q$ with $b_q$.

We therefore know that the Dirac equation, (1.3.A), will be valid for some wave function $\underline{\boldsymbol{\Phi}}_q\left(\phi_{1_q}, \phi_{2_q}\right)$ at the points, $b_q$, in $L$. This Dirac equation is

$$\left\{\underline{\mathbf{D}} - ie\underline{\mathbf{A}}_q^{\sim}\left(A_q^{\sim}, A_q^{\sim}\right)\right\}\underline{\boldsymbol{\Phi}}_q = \underline{\boldsymbol{\Phi}}_q \underline{\mathbf{M}}\left(m^{\sim}, -m^{\sim \ddagger}\right) \qquad (2.2.\text{G})$$

where $A_q^{\sim}$ and $\underline{\boldsymbol{\Phi}}_q$ vary with $q$, the location in $L$. We now have a Dirac equation, (2.2.G), that holds for particle $N^b$ at the points, $b_q$, when the particle interacts with the potential field, $iA_q^{\sim}$, described by the photon equation, (2.1.E), that also holds at $b_q$. Both these refer to $L$.

## 2.3  Local solution to the equations

We can bypass the photon and Dirac equations and consider the direct solution of equations (2.1.F), (2.1.G) and (2.2.B) as a set of simultaneous equations relating the values of our variables at point $q$. We show that the values produced are physically reasonable. We set

$$\rho = i\rho_q^{\sim}, \qquad A = iA_q^{\sim}, \qquad m = im^{\sim} \qquad (2.3.\text{A})$$

in order to clarify the behaviour of the variables more easily.

Substituting from equation (2.1.F) for $f_q$ and then equation (2.1.G) for $R_q$ into equation (2.2.B) we obtain

$$\frac{\rho^2}{de^2} - A^3\rho - m^2 dA^4 = 0 \qquad (2.3.\text{B})$$

where

$$d = \frac{3\pi}{n^2 h^2} \qquad (2.3.C)$$

giving

$$\rho = \frac{A^2 e^2 d}{2}\left( A \pm \sqrt{A^2 + \frac{4m^2}{e^2}} \right) \qquad (2.3.D)$$

The behaviour of this equation is satisfactory as we demonstrate. First, we treated $e$ and $m$ as finite and given and match these to their physical value. Here we are considering the isolated system of finite size at point $q$. We note that $m^2$, $e^2$ and $d$ in equation (2.3.D) are positive. We consider the positive square root. When $A$ is zero we must set $\rho$ to zero. As $A$ increases from zero so does $\rho$ without limit as $d\rho/dA$ is positive for all positive $A$. Thus we can always pick a positive value of $\rho$ for any given positive $A$ or vice versa. We consider the negative square root. Again, when $A$ is zero we must set $\rho$ to zero. As $A$ decreases from zero so does $\rho$ without limit as $d\rho/dA$ is still positive for all negative $A$. Therefore we can always pick a negative value of $\rho$ for any given negative $A$ or vice versa. After picking $\rho$ or $A$ such that $A$ and $\rho$ have the same sign, we may use equation (2.1.G) to provide a positive value of $R_q$ and then equation (2.1.F) to choose a value of $f_q$ with the opposite sign to $A$, as is necessary with our sign convention.

We see that we may follow this prescription whether the charges on $N^b$ and the $N_q^a$ have opposite signs or not. However, it is difficult to justify the Bohr equations, (1.3.K) and (1.3.L), if the signs are not opposite.

## 3.    QED FOR A SPACE OF ROUNDELS

### 3.1    The ensemble of roundels in $L$

We want to extend the solution above for the points, $b_q$, to the solution of the photon and Dirac equation over a space, where every point, $q$, has an associated particle, $N_q^a$. We model the potential using what we shall call *the ensembles of roundels in L*. These roundels are a subset of those centred on $q$ with radius $R_q$ that we have already described. In the rest frame of the particle $N_q^a$ at $q$, the roundel is either a circle for a pure state or a sphere for a superposition. We call this *the roundel frame*. In general, the roundel frame is different for each roundel. We pick a frame we shall call *the snapshot frame* for $L$, with Cartesian co-ordinates $(x_0', x_1', x_2', x_3')$, and consider a fixed time $x_0'$. We transform the roundels to the snapshot frame at this instant and insist they are sufficiently well-behaved to cover a non-zero area or volume as appropriate.

We call the subset of roundels we shall consider, $S_{q_j}$. We define the centre of the roundel, $S_{q_j}$, as the point $q_j$ and the radius in the roundel frame as $R_{q_j}$. The roundels, $S_{q_j}$, do not overlap, but touch at the boundaries in such a way that every point, $q$, in $L$ is no further than a distance $c R_{q_j}$ from a roundel boundary in every appropriate spatial direction: in a plane for a pure state and in three dimensions for a superposition. Here $R_{q_j}$ is the radius of any one chosen roundel and $c$ is any convenient constant.

We form a set, $F$, of the points on the boundaries of all the $S_{q_j}$. We would like to say without contradiction that the photon equation, (2.1.E), and Dirac equation, (2.2.G), hold on $F$. In order to be able to do this we assign points on the boundary of more than one roundel to the roundel, $S_{q_j}$, where point $q_j$ has the smallest $x_1'$ co-ordinate, or if these are the same, the smallest $x_2'$ co-ordinate, or if these are the same and we are therefore dealing with spheres in the roundel frame, the smallest $x_3'$ co-ordinate.

We divide $L$ into a set of regions, $j$. Each region consists of a finite set of roundels. The regions do not overlap, except on the region boundaries. Thus every point, $q$, in $L$ is part of one and only one region, unless it is on a region boundary. We specify that points on the boundaries of each region, $j$, shall also be on the boundary of some roundel, $S_{q_j}$. A point, $q$, on the boundary of a region will be assigned to a specific roundel. This roundel will have a centre in a unique region, $j$. The point $q$ is assigned to this region. *The ensemble of roundels, $E_j$, for $j$ is the union of all the $S_{q_j}$ whose centres are in $j$.*

We assume that all the $R_{q_j}$ and the $Z\left(R_{q_j}\right)$, where Z is a Lorentz transformation and rotation from the roundel to the snapshot frame, have the same order of magnitude. We also assume this for every instance of our other variables where necessary. We allow any $R_{q_j} \to 0$ while keeping equations (2.1.F), (2.1.G), (2.2.A) and (2.2.B) valid and also keeping the condition that every point in $L$ is no further from a roundel boundary than $cR_{q_j}$. Since the regions each contain a finite set of roundels, the volume or area, as appropriate, of the region also tends to zero. We also obtain $F \to L$. At the limit, the photon and Dirac equations, (2.1.E) and (2.2.G), apply to all points in $L$.

## 3.2   Limit behaviour for the roundels

We discuss how our variables behave when $R_{q_j} \to 0$, while equations (2.1.F), (2.1.G), (2.2.A) and (2.2.B) are kept valid. We use the roundel frame. We take it that both $m^{\~}$ and $e$ may be functions of $R_{q_j}$, as in renormalisation,[6] and replace them with the bare values, $m^{B\~}$ and $e^B$. We allow $R_{q_j} \to 0$ and set

$$m^{B\~} \sim \frac{1}{R_{q_j}} \qquad (3.2.A)$$

where ~ means *is commensurate with*. Equation (2.2.B) and relation (3.2.A) imply

$$e^B \sim \frac{1}{f_{q_j}} \qquad (3.2.B)$$

We call the particle, $N_q^a$, at the centre of roundel $S_{q_j}$, $N_{q_j}^a$. $f_{q_j}$ is the charge on particle $N_{q_j}^a$. The reason we choose this form of limit behaviour is that relation (3.2.B) means that the velocity, $-iv_q^{\~}$, in equation (2.2.A) stays finite as we scale the size of the roundel up or down by making $R_{q_j}$ larger or smaller, while keeping equation (2.2.B) valid by scaling $m^{B\sim}$. Therefore the period of the orbit, $t$, may be made as small as required to provide enough circuits of particle $N^b$ for us to treat particle $N^b$ as bound. This holds even though in fact we may be considering a portion of a wave function associated with some unbound state that we are sampling at intervals greater than $t$.

Consistency dictates another relation between the bare variables. Suppose we have an interaction between two charged particles, $N^a$ and $N^b$, forming a bound state. Let $N^a$ have bare charge $e^{B^a}$. We expect

$$e^{B^a} \sim e^B \qquad (3.2.C)$$

Before investigating what this implies, we must decide whether our two-body interaction is in a pure state or in a superposition. This is necessary because the model of the two-body interaction is different for each. We pick the former alternative.

Let the Bohr interaction be in the $M$-space $M^G$, where the superscript $G$ stands for global. We want to treat $N^a$ and $N^b$ symmetrically, and so we choose the centre of mass frame for the formation of the space $M^G$. The wave function for $N^b$ leads to a constant probability current for the location and motion of $N^b$ in $M^G$, as we discussed in part (1). This leads to a varying probability current in $L$, but the frame is constant. $N^b$ is forbidden by conservation of energy to appear anywhere but at the Bohr radius when $N^a$ is a point particle. We augment our previous model in part (1) by introducing a wave function for $N^a$ rather than assuming it is a point charge. Now $N^a$ is represented by a smeared-out wave function responsible for a probability current which generates the set of particles, $N^a_{q_j}$. $N^b$ can then appear at the boundary of the roundel, $S_{q_j}$, associated with particle $N^a_{q_j}$. By symmetry, what we have said about $N^b$ is also true of $N^a$. In particular, the probability current for $N^a$ has a constant frame. Since the rest frame of the $N^a_{q_j}$ depends upon this probability current, the frame is the same for all $N^a_{q_j}$. We create a single region and ensemble for the whole Bohr interaction.

The spatial part of the wave function of $N^b$ in $M^G$ and $L$ is two-dimensional, in the sense that the wave function depends only on the co-ordinates in the $(x_1, x_2)$ plane in $L$. Therefore the spatial part of the wave function of $N^b$ in the spaces, $M_q$, for roundel, $S_{q_j}$, must also depend on no more than the co-ordinates of the same plane. This holds because the union of such wave functions at the boundary of the roundels, $S_{q_j}$, in $L$ constitutes the complete wave function of $N^b$ in $L$. This means that for this choice of interaction the roundels are circles.

We normalise the interaction between $N^a$ and $N^b$ on a large but finite square with particle $N^a$ at the centre. This square remains of constant side $T$ as $R_{q_j}$ is varied. The number of local interactions taking place between the $N^a_{q_j}$ and $N^b$ is then

$$n^l \sim T^2 \big/ R_{q_j}^2 \qquad (3.2.\text{D})$$

where we have assumed we can approximate the ensemble by a two-dimensional lattice of minimum interval $2R_{q_j}$ to the correct order of magnitude. Since we have

$$e^{B^a} = \sum_{q_j} f_{q_j} \sim n^l f_{q_j} \qquad (3.2.\text{E})$$

we obtain from relation (3.2.D)

$$e^{B^a} \sim \frac{f_{q_j}}{R_{q_j}^2} \qquad (3.2.\text{F})$$

From relation (3.2.B) and our assumption that $e^B$ and $e^{B^a}$ show similar behaviour, we deduce that

$$e^B \sim \frac{1}{R_{q_j}}, \qquad \mathrm{e}^{B^a} \sim \frac{1}{R_{q_j}} \qquad (3.2.\text{G})$$

We suppose the renormalisation of the bare charge does not depend on the type of application, provided the infinitesimal two-body interaction remains a pure state, and return to a more general description. From equations (3.2.B) and (3.2.G) we obtain

$$f_{q_j} \sim R_{q_j} \qquad (3.2.\text{H})$$

Equation (2.1.F) and relation (3.2.H) lead to

$$A_{q_j}^{\tilde{\ }} \sim 1 \qquad (3.2.\text{I})$$

Where $iA_{q_j}^{\sim}$ is the potential on the boundary of roundel $S_{q_j}$. From equation (2.1.G) and relation (3.2.I) we obtain

$$\rho_{q_j}^{B\sim} \sim \frac{1}{R_{q_j}^2} \qquad (3.2.J)$$

where $i\rho_{q_j}^{B\sim}$ is the charge density for roundel $S_{q_j}$. We see that $\rho_{q_j}^{B\sim} \to \infty$ as $R_{q_j} \to 0$. This means the differentials of $A_{q_j}^{\sim}$ are not well-behaved in the photon equation, (2.1.E). The ensemble of roundels, as we have constrained it so far, permits discontinuous behaviour. When we put the photon and Dirac equation on a lattice we shall introduce different behaviour in the limit when the lattice interval goes to zero, that will enable us to keep $\rho_{q_j}^{B\sim}$ finite.

## 4.  QED FOR A LATTICE

### 4.1  Mapping from the roundels to a lattice

We show that the description in terms of roundels can also be put into the better-known language of lattice gauge theory.[9, 11-13] We define hypercubic lattice, $L'$, that covers $L$. Lattice $L'$ has lattice points, $q'_{j'}$, with a constant minimum separation of $2a$. We choose the snapshot frame and arrange that the axes of our co-ordinate system, $(x'_0, x'_1, x'_2, x'_3)$, lie along the lattice vectors where the lattice points have an interval of $2a$. We use a procedure to pick each lattice point, $q'_{j'}$, in turn. Suppose we have picked lattice point $k'_{j'}$. We form a region, $j'$, from lattice points $q'_{j'} = k'_{j'}$, $k'^{-\mu}_{j'}$ and $k'^{+\mu}_{j'}$, where $k'^{-\mu}_{j'}$ and $k'^{+\mu}_{j'}$ are the nearest lattice point in the direction of the $-x'_{\mu}$ and $+x'_{\mu}$ axis, respectively.

We find the region, $j$, in which $k'_{j'}$ lies and pick the roundel, $S_{q_j}$, whose centre, $q_j$, is nearest to $k'_{j'}$. We call this roundel $S_{k_j}$, the radius $R_{k_j}$ and the centre $k_j$. We call the roundel frame for $S_{k_j}$ *the compromise frame* because we approximate by applying it as the frame for the whole ensemble in region $j$. We take the radii of the roundels in $j$ as $R_{k_j}$. We take the charge density of the region $j$ as $i\rho^{B\sim}_{k_j}$, absorbing a factor due to the difference between the volume covered by the cube of side $2R_{k_j}$ and sphere of radius $R_{k_j}$ in the units.

We assume that the centres of the roundels, $q_j$, in the ensemble of roundels, $E_j$, form an approximate lattice, which we will call $L_k$. We need to make an assumption about how the roundels are packed for an explicit calculation. We shall assume that we have a superposition state where the roundels are spheres and our lattice, $L_k$, is cubic with a minimum interval between lattice points of $2R_{k_j}$. Other assumptions about the type of lattice will lead to similar conclusions.

Let $\underline{\boldsymbol{\Phi}}_{q_j}$ be the wave function on the boundary of roundel $S_{q_j}$. The form of the variables associated with $L_k$, that is, $\rho^{B\sim}_{k_j}$, $A^{\sim}_{q_j}$ and $\underline{\boldsymbol{\Phi}}_{q_j}$, do not change as time passes for the interval of interest and so we can turn $L_k$ into a hypercubic lattice by entering extra lattice points, $q_j$, along lines parallel to the time axis, $x_0$. One such line passes through every old lattice point and the extra lattice points are added at intervals of $2R_{k_j}$. Each new lattice point is associated with copies of the same variables, $\rho_{k_j}^{B\sim}$, $A_{q_j}^{-}$ and $\underline{\Phi}_{q_j}$, referring to successive times. $L_k$ becomes a hypercubic lattice.

The compromise frame is addressed by the Cartesian co-ordinates $(x_0, x_1, x_2, x_3)$. We arrange that the axes should lie along the $L_k$ lattice vectors where the lattice points have an interval of $2R_{k_j}$. In the limit as $R_{k_j} \to 0$, $L_k$ for region $j$ becomes an exact lattice of minimum interval $2R_{k_j}$ and with charge density $i\rho_{k_j}^{B\sim}$. We insist that the lattice points in $L_k$, $q_j = k_j^{-\mu}$ and $k_j^{+\mu}$, where $k_j^{-\mu}$ and $k_j^{+\mu}$ are the nearest lattice point in the direction of the $-x_\mu$ and $+x_\mu$ axis, respectively, should be part of region $j$.

We map the lattice points in region $j$ in the compromise frame onto lattice points in region $j'$ in the snapshot frame with the bijection

$$k_{j'}', k_{j'}'^{-\mu}, k_{j'}'^{+\mu} = \hat{Z}\big(k_j, k_j^{-\mu}, k_j^{+\mu}\big) \qquad (4.1.A)$$

where $\hat{Z}$ is a transformation that consists of a translation that maps $k_j$ to $k_{j'}'$, a scaling and then, with $k_{j'}'$ as the origin, the same Lorentz transformation and rotation, Z, as we used in section 3.1. This maps $k_j^{-\mu}$ onto $k_{j'}'^{-\mu}$ and $k_j^{+\mu}$ onto $k_{j'}'^{+\mu}$. We place the same particles, $N_{q_j}^a$, associated with the same charges, $f_{q_j}$, for the lattice points $q_j = k_j$, $k_j^{-\mu}$ and $k_j^{+\mu}$ in $L_k$, on the lattice points, $q_{j'} = k_{j'}'$, $k_{j'}'^{-\mu}$ and $k_{j'}'^{+\mu}$, respectively, in $L'$. The $N_{q_j}^a$ on the lattice points $k_{j'}'$, $k_{j'}'^{-\mu}$ and $k_{j'}'^{+\mu}$ remain at rest in the compromise rather than the snapshot frame.

## 4.2 Values for lattice variables

We suppress the subscript *j* on the understanding that all our points

are either lattice points in one of the lattices $L_k$ or $L'$ except those

specifically indicated. We write our variables for *j* in invariant notation

for point *k* but do not otherwise alter them

$$\rho_k^{B\sim} \to \mathbf{J}_k^{\sim}, \qquad\qquad A_q^{\sim} \to \mathbf{A}_k^{\sim}, \qquad\qquad (4.2.\text{A})$$
$$m^{B\sim} \to \mathbf{M}^{\hat{B}\sim}, \qquad\qquad \underline{\Phi}_q \to \underline{\Phi}_{\hat{k}}\left(\phi_{1_k}, \phi_{2_{\hat{k}}}\right)$$

Let the charge density current for *j'* at point *k'* be $\mathbf{J}_{k'}^{\sim}$. We have

$$\left|\mathbf{J}_{k'}^{\sim}\right| = \frac{f_k}{a^3} \qquad\qquad (4.2.\text{B})$$

where $f_k$ is the charge on particle $N_k^a$ at lattice points *k* and *k'*. We obtain

from equations (4.2.A), (4.2.B), equations corresponding to (2.1.F) and

(2.1.G) for $\left|\mathbf{J}_k^{\sim}\right|$, and the definition of Z

$$\mathbf{J}_{k'}\left(\mathbf{J}_k^{\sim}, \mathbf{J}_{k'}^{\sim\ddagger}\right) = \frac{R_k^3}{a^3} Z\left\{\mathbf{J}_k\left(\mathbf{J}_k^{\sim}, \mathbf{J}_k^{\sim\ddagger}\right)\right\} \qquad\qquad (4.2.\text{C})$$

Let the point on the boundary of roundel $S_k$, midway between lattice point $k$ and $k^{+\mu}$, in $L_k$ be called $k^{+\mu}$. The potential term, $\mathbf{A}_k^{\sim}$, at $k^{+\mu}$ due to the charge, $f_k$, at $k$ is $if_k/R_k$ in the compromise frame, from equation (2.1.F). Let the point midway between lattice locations $k'$ and $k'^{+\mu}$ in the lattice $L'$ be called $k'^{+\mu}$. Let the potential term at $k'^{+\mu}$ be $\mathbf{A}_{k'}^{\sim}$ in the snapshot frame. Since this potential term is $if_k/a$ in the compromise frame,

$$\left|\mathbf{A}_{k'}^{\sim}\right| = f_k/a. \tag{4.2.D}$$

We obtain from equation (4.2.D) and the definition of Z,

$$\underline{\mathbf{A}}_{k'}\left(\mathbf{A}_{k'}^{\sim}, \mathbf{A}_{k'}^{\sim \ddagger}\right) = \frac{R_k}{a} Z\left\{\underline{\mathbf{A}}_k\left(\mathbf{A}_k^{\sim}, \mathbf{A}_k^{\sim \ddagger}\right)\right\} \tag{4.2.E}$$

Let the point on the boundary of roundel $S_q$, where $q = k^{-\mu}$, and midway between a lattice point $k$ and $k^{-\mu}$ in $L_k$, be called $k^{-\mu}$. Let the potential term at $k^{-\mu}$ be $\mathbf{A}_{k^{-\mu}}^{\sim}$. We set the partial differential in lattice $L_k$ to the discrete differential on the understanding that eventually we will set $R_k \to 0$,

$$\frac{\partial \mathbf{A}_k^{\sim}}{\partial x_\mu} = \frac{\mathbf{A}_k^{\sim} - \mathbf{A}_{k^{-\mu}}^{\sim}}{2R_k} \tag{4.2.F}$$

Let the point midway between a lattice point $k'$ and $k'^{-\mu}$ in $L'$ be called $k'^{-\mu}$. Let the potential term at $k'^{-\mu}$ be $\mathbf{A}_{k'^{-\mu}}^{\sim}$. We set the partial differential in lattice $L'$ to the discrete differential on the understanding that eventually we will set $a \to 0$, and obtain

$$\frac{\partial \mathbf{A}_{\tilde{k}'}}{\partial x'_\mu} = \frac{\mathbf{A}_{\tilde{k}'} - \mathbf{A}_{\tilde{k}'-\mu}}{2a} \tag{4.2.G}$$

We use equations (4.2.E), (4.2.F) and (4.2.G) to find

$$\frac{\partial \underline{\mathbf{A}}_{\tilde{k}'}\left(\mathbf{A}_{\tilde{k}'}^{\sim}, \mathbf{A}_{\tilde{k}'}^{\sim \ddagger}\right)}{\partial x'_\mu} = \frac{R_k^2}{a^2} Z\left(\frac{\partial \underline{\mathbf{A}}_k\left(\mathbf{A}_k^{\sim}, \mathbf{A}_k^{\sim \ddagger}\right)}{\partial x_\mu}\right) \tag{4.2.H}$$

Let $\mathbf{i}'_\mu$ be the equivalent, for the snapshot frame, of $\mathbf{i}_\mu$ for the compromise frame, where $\mathbf{i}_\mu$ is as defined in part (1). Multiplying both sides of equation (4.2.H) by $\underline{\mathbf{i}}'_\mu\left(\mathbf{i}'_\mu, \mathbf{i}'^{\ddagger}_\mu\right)$ and summing over $\mu$, we obtain

$$\underline{\mathbf{D}}_{\tilde{k}'}\left(\mathbf{D}', \mathbf{D}'^{\ddagger}\right)\underline{\mathbf{A}}_{\tilde{k}'} = \frac{R_k^2}{a^2} \sum_\mu \left\{ \underline{\mathbf{i}}'_\mu \, Z\left(\frac{\partial \underline{\mathbf{A}}_k}{\partial x_\mu}\right) \right\} \tag{4.2.I}$$

where

$$\mathbf{D}' = \mathbf{i}'_0 \frac{\partial}{\partial x'_0} + \mathbf{i}'_1 \frac{\partial}{\partial x'_1} + \mathbf{i}'_2 \frac{\partial}{\partial x'_2} + \mathbf{i}'_3 \frac{\partial}{\partial x'_3} \tag{4.2.J}$$

On the right-hand-side of equation (4.2.I) we are transforming the components, $\partial \underline{\mathbf{A}}_k / \partial x_\mu$, while leaving the matrices, $\underline{\mathbf{i}}'_\mu$, the same. We swap to transforming the matrices rather than the components, as discussed by Bell et al.,[2] obtaining

$$\underline{\mathbf{D}}_{\tilde{k}'}\left(\mathbf{D}', \mathbf{D}'^{\ddagger}\right)\underline{\mathbf{A}}_{\tilde{k}'} = \frac{R_k^2}{a^2} Z\left\{\underline{\mathbf{D}}_k\left(\mathbf{D}, \mathbf{D}^{\ddagger}\right)\underline{\mathbf{A}}_k\right\} \tag{4.2.K}$$

where $\mathbf{D}$ is as defined in part (1). Equations (4.2.E) and (4.2.K) imply that

$$\underline{\mathbf{D}}_{k'} = \frac{R_k}{a} \, \mathsf{Z}(\underline{\mathbf{D}}_k) \tag{4.2.L}$$

## 4.3   Lattice version of the photon and Dirac equations

The photon equation, (2.1.E), in the invariant form of equation (1.3.B) for the lattice $L_k$ is

$$\underline{\mathbf{D}}_k \underline{\mathbf{D}}_k \underline{\mathbf{A}}_k = \underline{\mathbf{J}}_k \left( \mathbf{J}_k^{\sim}, \mathbf{J}_k^{-\frac{\sim}{\sim}} \right) \tag{4.3.A}$$

We multiply each side of equation (4.3.A) by $R_k^3 / a^3$, transform by $\mathsf{Z}$ and obtain from equations (4.2.C), (4.2.L), (4.2.E)

$$\underline{\mathbf{D}}_{k'} \underline{\mathbf{D}}_{k'} \underline{\mathbf{A}}_{k'} = \underline{\mathbf{J}}_{k'} \tag{4.3.B}$$

which is the photon equation on the lattice $L'$.

The Dirac equation, (2.2.G), for the lattice $L_k$, in the invariant form of equation (1.3.A), is

$$\left( \underline{\mathbf{D}}_k - ie^{\,B} \underline{\mathbf{A}}_k^{\sim} \right) \underline{\boldsymbol{\Phi}}_{\hat{k}} = \underline{\boldsymbol{\Phi}}_{\hat{k}} \, \underline{\mathbf{M}}^{\hat{B}\sim} \left( \mathbf{M}^{\hat{B}\sim}, -\mathbf{M}^{\hat{B}\sim\frac{\sim}{\sim}} \right) \tag{4.3.C}$$

from equation (4.2.A) and the definition of $\underline{\mathbf{D}}_k$ in equation (4.2.K). An approach similar to that taken for the photon equation encounters a difficulty. Lattice $L_k$ has a local scaling factor, $R_k$, which varies with the region, $j$, while $\mathbf{M}^{\hat{B}\sim}$ is a global variable. At this point we are forced to renormalise the mass of $N^b$ differently for each region, $j$, according to

$$\mathbf{M}_k^{\hat{B}\sim} = \frac{a}{R_k} \mathbf{M}^{\hat{B}\sim} \tag{4.3.D}$$

where $\mathbf{M}_k^{\hat{B}\sim}$ is the energy-momentum term for each region $j$ and the corresponding lattice $L_k$ in the compromise frame. The value of $f_q$ in equation (2.2.B), where $q = k$, remains unaltered. We also define

$$\mathbf{M}^{B\sim} = Z\left(\mathbf{M}^{\hat{B}\sim}\right) \tag{4.3.E}$$

where $\mathbf{M}^{B\sim}$ is the global energy-momentum term for the lattice $L'$ in the snapshot frame. Replacing $\mathbf{M}^{\hat{B}\sim}$ with $\mathbf{M}_k^{\hat{B}\sim}$ in equation (4.3.C), we obtain

$$\left(\underline{\mathbf{D}}_k - ie^B \underline{\mathbf{A}}_k^\sim\right)\underline{\boldsymbol{\Phi}}_k\left(\boldsymbol{\phi}_{1_k}, \boldsymbol{\phi}_{2_k}\right) \tag{4.3.F}$$
$$= \underline{\boldsymbol{\Phi}}_k\left(\boldsymbol{\phi}_{1_k}, \boldsymbol{\phi}_{2_k}\right)\underline{\mathbf{M}}_k^{\hat{B}\sim}\left(\mathbf{M}_k^{\hat{B}\sim}, -\mathbf{M}_k^{\hat{B}\sim\ddagger}\right)$$

where $\underline{\boldsymbol{\Phi}}_k$ is the wave function associated with the new mass term $\underline{\mathbf{M}}_k^{\hat{B}\sim}$. We multiply each side of equation (4.3.F) by $R_k/a$, transform by $Z$ and obtain from equations (4.2.E), (4.2.L), (4.3.D) and (4.3.E)

$$\left(\underline{\mathbf{D}}_{k'} - ie^B \underline{\mathbf{A}}_{k'}^\sim\right)\underline{\boldsymbol{\Phi}}_{k'}\left(\boldsymbol{\phi}_{1_{k'}}, \boldsymbol{\phi}_{2_{k'}}\right) = \tag{4.3.G}$$
$$\underline{\boldsymbol{\Phi}}_{k'}\left(\boldsymbol{\phi}_{1_{k'}}, \boldsymbol{\phi}_{2_{k'}}\right)\underline{\mathbf{M}}^{B\sim}\left(\mathbf{M}^{B\sim}, \mathbf{M}^{B\sim\ddagger}\right)$$

where $\underline{\boldsymbol{\Phi}}_{k'}$ is the wave function for lattice $L'$.

## 4.4  Limit behaviour for the lattice

We investigate the bare values of the mass and charge on lattice $L'$. We revise equation (2.1.G), obtaining

$$\left|\mathbf{A}_{k'}^\sim\right| = \frac{4\pi a^2 \left|\mathbf{J}_{k'}^\sim\right|}{3} \tag{4.4.A}$$

We shall require the differentials of $\mathbf{A}_{\tilde{k}'}$ to be well-behaved in the photon equation, (4.3.B). This means that

$$\left| \mathbf{J}_{\tilde{k}'} \right| \sim 1 \qquad (4.4.B)$$

Equation (4.4.A) then leads to the limit behaviour, when $a \to 0$,

$$\left| \mathbf{A}_{\tilde{k}'} \right| \sim a^2 \qquad (4.4.C)$$

Substituting from equation (4.4.C) into equation (4.2.D), we obtain

$$f_k \sim a^3 \qquad (4.4.D)$$

We use an argument similar to that used in section 3.2 to obtain the behaviour of $e^B$ as $R_q \to 0$. We consider an interaction between the two particles, $N^a$ with charge $e^{B^a}$ and $N^b$ with charge $e^B$ as before. However, we intend to use equation (4.3.G) for a superposition rather than a pure state. We therefore have a Bohr interaction for $N^b$ and $N^B$ which is also a superposition. As we discussed in part (1), this superposition is still an eigenstate of $\mathbf{m}_{x_3}$ with a constant frame for the probability density current in $L$. We introduce a wave function for $N^a$ rather than considering it a point particle, and by symmetry we expect a constant frame for the probability density current in $L$. Since the rest frame of the particles $N_k^a$ placed on the lattice points of $L'$ depends upon the probability current for $N^a$, the frame is the same for all $N_k^a$. We may therefore choose to make this the snapshot frame.

We remark on the ensemble of roundels. We see by comparing this discussion with our similar remarks in section 3.2 that there is parity between superpositions and pure states for the ensemble of roundels even before the Bohr interaction is put on lattice $L'$. For both states we may create a single region and ensemble for the whole Bohr interaction.

Returning to lattice $L'$, we normalise the interaction between $N^a$ and $N^b$ on a large but finite cube with particle $N^a$ at the centre. This cube remains of constant side $T$ as $a$ is varied. The number of local interactions taking place between the $N^a_k$ and $N^b$ is then

$$n^{l'} \sim T^3 \big/ a^3 \qquad (4.4.\text{E})$$

Since we have

$$e^{B^a} = \sum_{k'} f_k \sim n^{l'} f_k \qquad (4.4.\text{F})$$

we obtain from relation (4.4.E)

$$e^{B^a} \sim \frac{f_k}{a^3} \qquad (4.4.\text{G})$$

As before $e^B \sim e^{B^a}$ and from relation (4.4.D)

$$e^{B^a} \sim 1, \qquad e^B \sim 1 \qquad (4.4.\text{H})$$

We find that the charge does not need renormalisation when we put the photon and Dirac equation on a lattice by this route.

We suppose the renormalisation of the bare charge does not depend on the type of application, provided the infinitesimal two-body interaction remains a superposition, and return to a more general description. We may restate equation (2.2.B) for lattice $L_k$, point $k$ as

$$\left| \mathbf{M}^{B\sim} \right| R_k = -i \frac{n^2 h^2 \sqrt{1 - \frac{4\pi^2 e^{B2} f_k^2}{n^2 h^2}}}{4\pi^2 e^B f_k} \tag{4.4.I}$$

from equations (4.2.A) and (4.3.E). From equations (4.4.D) and (4.4.H) we obtain

$$\left| \mathbf{M}^{B\sim} \right| \sim \frac{1}{a^3 R_k} \tag{4.4.J}$$

We must have $R_k \to 0$ as $a \to 0$ because our method in section 4 depends on it. We may conveniently set $R_k = a^p$, $p > 0$, in which case

$$\left| \mathbf{M}^{B\sim} \right| \sim \frac{1}{a^{3+p}} \tag{4.4.K}$$

We see a power law relation typical of lattice behaviour at criticality.

## 5.    DISCUSSION

We see that the quantum theory is QED in the instance where the two-body interaction in the $M_q$ is given by the Bohr equations, (2.2.A) and (2.2.B), and provided the charge density current and potential field are sufficiently well-behaved.

We have only considered attractive interactions. Once we have proved that the equations hold for these we may use charge conjugation on the versatile Dirac equation[2] to prove that the Dirac equation continues to hold when the sign on particle $N^b$ is changed and the interaction becomes repulsive.

We may consider the inverse of our argument above in part (2) of this paper and ask whether the interaction of a potential field obeying the photon equation and a particle obeying the Dirac equation in $L$ may be pictured instead as a sea of infinitesimal Bohr interactions in the $M_q$. There is one difficulty. The Dirac equation for the two-body interaction may lead to a solution where the energy levels exhibit fine structure.[6] Bell et al.[4] and Bell and Diaz[5] extend the parameter, $R$, by adding a second similar parameter, $\hat{R}$. In addition to a curve specified by $R$, which maps to the $(x_1, x_2)$ plane in $L$, the definition of $M_q$ includes a curve, specified by $\hat{R}$, that maps to the $(x_0, x_3)$ plane. Bell et al. then consider a similar model for an attractive two-body interaction but this time with a circular orbit with two circular components, the first mapping to the $(x_1, x_2)$ plane and the second to the $(x_0, x_3)$ plane in $L$. Both components obey the Bohr equations. Difficulties with time-like loops are avoided by taking a co-ordinate in the $(x_1, x_2)$ plane as the time co-ordinate when applying the Bohr equations to motion in the $(x_0, x_3)$ plane. This accounts for all the possible eigenvalues of the bound system. Our argument in this paper can then be applied to the circular components in both planes of the orbit and the inverse holds.

We have constrained the same infinities that require renormalisation to be used to the mass of the particle obeying the Dirac equation in our lattice rendition of QED, but we have not banished them. However, it is also interesting to consider more rather than less unruly behaviour. This is permitted if we take the ensemble of roundels as the prior descriptor of our interaction. For example, adjacent points might have different levels of excitation with *n* varying. Provided *n* was sufficiently well-behaved for an average, $\bar{n}$, to be validly calculated for small regions, then the method we used here would stand. It would also be possible to look at macroscopic behaviour using different methods for generalising to physical distances. Instead of modelling the ensemble of roundels using a lattice, discrete fractals could provide a model instead.[1, 10] With such an approach it might not be necessary to avoid discontinuous change. It would then appear possible that this model of QED could be used to describe multiple systems with different local quantum states although the model uses no more than a particle description.

## ACKNOWLEDGEMENTS

One of us (Bell) would like to acknowledge the assistance of E.A.E. Bell.